# DWT-GBT-SVD-based Robust Speech Steganography

Noshin Amiri[1], Iman Naderi[2].
[1] Islamic Azad University, Doroud Branch, Lorestan, Iran, amirinoshin20@gmail.com
[2] Islamic Azad University, Boroujerd Branch, Lorestan, Iran

# Abstract

Steganography is a method that can improve network security and make communications safer. In this method, a secret message is hidden in content like audio signals that should not be perceptible by listening to the audio or seeing the signal waves. Also, it should be robust against different common attacks such as noise and compression. In this paper, we propose a new speech steganography method based on a combination of Discrete Wavelet Transform, Graph-based Transform, and Singular Value Decomposition (SVD). In this method, we first find voiced frames based on energy and zero-crossing counts of the frames and then embed a binary message into voiced frames. Experimental results on the NOIZEUS database show that the proposed method is imperceptible and also robust against Gaussian noise, re-sampling, re-quantization, high pass filter, and low pass filter. Also, it is robust against MP3 compression and scaling for watermarking applications.

## 1. Introduction

Steganography is a type of communication in which a very important and confidential message is hidden in content such as video, audio, or image. The main purpose of Steganography is to create a very secure and completely indistinguishable communication system. So that there is not even any doubt about sending confidential information. If a Steganography method causes a person to suspect a hidden message in the content, then that method will be rejected. Therefore, one of the most important factors in Steganography is its Imperceptibility or invisibility.

There are different types of Steganography. A confidential message can be embedded in a text, an image, an audio file, or a video. This article introduces a method of audio Steganography. Audio steganography is the hiding of a confidential message in a cover audio signal so that it is both robust and imperceptible. Security and robustness are two vital features of audio Steganography.

One of the important uses of audio steganography is in battlefield communication. In computer-based steganography systems, confidential messages are usually embedded in a digital audio signal as a sequence of binary numbers. Existing systems are able to embed messages in WAV, AU, and MP3 files.

The basic model of audio steganography consists of three parts: cover (audio file), message, and password key. The cover is also called the carrier because the message is sent through it. The message is that the sender hopes to remain confidential until it reaches the intended recipient. The message can be text, image, audio, or any other data. The password key, also called the Stego-key, is part of the message extraction algorithm that only the recipient has, and without the key, it is impossible to extract a confidential message from the cover.

Hiding information involves two main steps:
1- Determining redundant bits in the cover signal.
2- Placing the secret message bits instead of the extra bits of the cover file.

Additional bits are bits that do not change the quality of the cover signal. After embedding the message, the resulting audio signal is called a stego audio.

Audio steganography is more challenging than other types of steganography. One of these challenges is that with the compression of audio, a large part of the information of the audio is lost, and then the information of the confidential message is also lost. Audio transmissions may also have fading or noise. Also, if someone suspects the existence of confidential information in the voice, even if they cannot access the content of the message, they can destroy the message in the audio signal with intentional attacks on the stego audio. Another case is that the range of the Human Auditory System (HAS) is much larger than that of the Human Vision System (HVS), and when a message is embedded in audio, HAS is able to detect it as noise. Even in many cases where HAS is unable to detect, changes in the audio may be visible by drawing a temporal chart or a spectrogram. Therefore, it is necessary that the audio steganography method, while being robust against various attacks, is also Imperceptible.

In this paper, we propose a new robust audio steganography method based on Discrete Wavelet Transform (DWT) and Graph-based Transform (GBT). GBT is a new transformation that is using today in different applications such as audio compression [1] and audio watermarking [2]. GBT acts like a Discrete Cosine Transform (DCT) which de-correlates the input signal. But in some studies, it has been shown that GBT outperforms DCT.

The rest of the paper is structured as follows:

In section 2, some related studies have been introduced. In section 3, DWT and GBT have been explained. In section 4, the proposed steganography method has been described. In section 5, experimental results have been reported and in section 6, the conclusion has been provided.

## 2. Related Works

In [3], after introducing different methods of audio steganography, two methods of steganography based on DCT and DWT have been introduced.

In [4], a method based on DCT for audio steganography is presented, in which voiced and unvoiced characteristics are used. The basic idea is that message bits change the amplitude of the DCT coefficients, but the changes are different for the voiced and unvoiced parts of the audio signal. The results show that this method is able to provide a signal-to-noise ratio (SNR) of 43.9 dB and a payload capacity of 1.08 Kbps.

In [5], a DCT-based method and Singular Value Decomposition (SVD) were used for Speech signal steganography. In this method, only voiced frames are used to embed the message, which increases the imperceptibility of the steganography.

In [6], Generative Adversarial Networks (GANs) are used for audio steganography. In this method, three networks have used that work together at the same time. The first network is responsible for embedding the message in the signal, the second network is responsible for extracting the message from the stego signal, and the third network is a discriminative network that determines which audio frames should contain message bits.

In [7], Adaptive Huffman Code Mapping is used to hide information in the audio signal. In this method, psychoacoustic models and characteristics of the human auditory system (HAS) are used to increase the imperceptibility of steganography. Adaptive Hoffman code in this method has increased the secure payload capacity.

In [8], a random model has used to embed data in audio signals. In this way, message bits can be placed in any position of the signal. Although this approach increases computational complexity, it greatly increases the level of security.

## 3. Preliminaries

In the proposed method, we use to embed the message bits only in the voiced frames of the cover audio signal. So, we use an algorithm to determine the voiced frames. We use Graph-based Transform, Discrete Wavelet Transform, and Singular Value Decomposition as well. So, in this section, we explain these four main parts of the proposed method.

### 3-1- Voiced Frames Detection

In general, an audio signal can be divided into three parts: voiced, un-voiced, and silent. The voice from trapped air produced by the mouth of the larynx (glottis) is called voiced. This sound is created when vocal cords vibrate. An un-voiced sound is created when the relaxed vocal cords allow the trapped air to be released. The voiced part of a speech signal contains high-amplitude and low-frequency components. While the unvoiced part contains the low-amplitude and high-frequency components. The part of silence that is usually present at the beginning and end of speech does not have significant frequency and amplitude components [9].

With two values of Zero Crossing Count (ZCC) and Short-time Energy (STE), voiced and unvoiced frames can be separated. Eq. 1 describes how to separate frames:

$$frame = \begin{cases} Voiced & , \quad if\ ZCC\ is\ LOW\ and\ STE\ is\ HIGH \\ Unvoiced & , \quad \quad \quad \quad \quad \quad \quad \quad \quad \quad Otherwise. \end{cases} \quad (1)$$

The value of ZCC is the number of times the signal passes through zero over time, which is directly related to the signal frequency. Therefore, if the ZCC is high, it means that the frequency is also high and the probability that the frame is unvoiced is higher. The ZCC value is calculated based on Eq. 2:

$$ZCC = \frac{1}{2}\sum_{n=1}^{N}|sign(f[n]) - sign(f[n-1])| \quad (2)$$

Where $f[n]$ is the $n^{th}$ sample of the frame, $f$, and $N$ is the number of total samples of $f$. The sign value is also calculated as follows:

$$sign(f[n]) = \begin{cases} +1 & , \quad if\ f[n] > 0 \\ -1 & , \quad Otherwise. \end{cases} \quad (3)$$

The energy of each frame is the sum of the squares of the samples in that frame. But it should be noted that framing is done using the Hamming window. Therefore, the value of each sample should be calculated by multiplying it in the weight of the corresponding hamming window. The weight of each sample from the Heming window for a window with length $L$ is calculated as Eq. 4:

$$w[n] = \begin{cases} 0.54 - 0.46\cos\frac{2\pi n}{L-1} & for\ 1 \leq n \leq L \\ 0 & Otherwise. \end{cases} \quad (4)$$

With Heming's weights, the short-time energy of a frame is calculated as the equation 5:

$$STE = \sum(f[n]w[m-n])^2 \quad (5)$$

### 3-2- Graph-based Transform (GBT)

Given a block of an audio signal with a frame size of $N$ samples, we can create a graph $G=\{V,E,s\}$ where $V$ and $E$ are the vertices and edges of the graph, and $s \epsilon \mathbb{R}^{N \times 1}$ is an audio signal for which the graph matrix is defined as $K \epsilon \mathbb{R}^{N \times N}$. For this graph, the adjacency matrix A, elements are obtained as

$$A_{ij} = \begin{cases} a_{ij}, & if\ (i.j) \in E \\ 0, & otherwise \end{cases} \quad (6)$$

Where $a_{ij}$ is the weight of the edge between $i$ and $j$ in the graph. The degree matrix $D \epsilon \mathbb{R}^{N \times N}$ is a diagonal matrix, for which the elements are defined as follows,

$$D_{ij} = \begin{cases} \sum a_{ij}, & if\ i = j \\ 0, & otherwise \end{cases} \quad (7)$$

Then, the Graph-Laplacian Matrix $L$ would be defined as,

$$L = D - A \quad (8)$$

Where the operator $L$ is also known as *Kirchhoff* operator, as a tribute to Gustav Kirchhoff for his achievements on electrical networks. Kirchhoff referred to the (weighted) adjacency matrix $A$ as the *conductance* matrix.

The matrix L would be a real symmetric one and based upon the spectral theory, the eigenvalue decomposition (EVD) of this matrix would lead to a set of real non-negative eigenvalues, denoted by $\Lambda = \{\lambda_1, \ldots, \lambda_N\}$, and a set of corresponding independent (hence, orthogonal) eigenvectors denoted by $V = \{v_1, \ldots, v_N\}$, derived as,

$$L = V \Lambda V^T \quad (9)$$

Then we can use these orthogonal eigenvectors to de-correlate the signal defined on the graph, i.e.,

$$c = V^T s \quad (10)$$

Where $c \epsilon \mathbb{R}^{N \times 1}$ is the approximate sparse transform coefficient matrix [10].

### 3-3- Discrete Wavelet Transform

Wavelet Transform (WT) analyses a signal in different frequencies with different resolutions. It can show a signal with different components which are known as wavelets. In general, WT provides a high time resolution with low-frequency resolution and a low time resolution with high-frequency resolution. An audio signal in the digital domain is a discrete signal. So, in this paper, we use the discrete version of WT which called Discrete Wavelet Transform (DWT).

The main idea of the DWT is to divide an audio signal into a high-frequency part and a low-frequency part. So, the signal needs to pass through a high-pass filter and a low-pass filter.

The high-frequency and the low-frequency components of an input signal x can be determined as follows:

$$y_{high}[k] = \sum_n x[n]\, g[2k - n] \quad (11)$$

$$y_{low}[k] = \sum_n x[n]\, h[2k - n] \quad (12)$$

$y_{low}$, which is calculated by a high pass filter, is called a detailed signal, and $y_{high}$ is called approximation signal. To maintain in a constant length, a down-sampling with a stride of 2 is used for each filter. Fig. 1 shows the DWT algorithm.

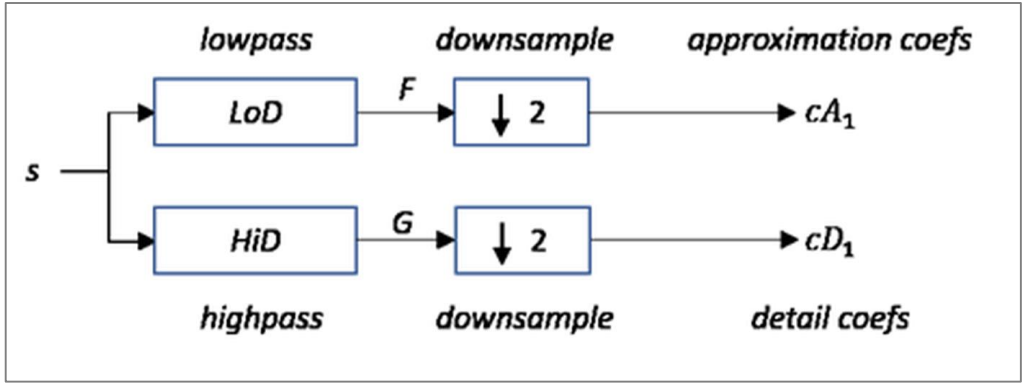

*Figure 1. DWT algorithm*

### 3-4 Singular Value Decomposition

Singular Value Decomposition (SVD) is a powerful mathematical tool that provides three matrices from one given matrix as follows:

$$A = USV^T \qquad (13)$$

Where $U$ is a $m \times m$ unitary matrix with real numbers. i.e. $UU^t = I$.
$S$ is a $m \times n$ diagonal matrix that includes non-zero real number in its diagonal as singular values.
$V$ is a $n \times n$ unitary matrix with real numbers.
In this paper, we use the $S$ matrix for embedding the message bits. In other words, we add the message bits to the biggest singular value in the $S$ matrix.

## 4. Proposed method

The proposed speech steganography method has two main parts: embedding and extraction of a secret message. The embedding part is done as follows:

1- Dividing the cover speech signal into different frames in equal lengths.
2- Detecting voiced frames between all existing frames.
3- Selecting $N$ voiced frames which have lower ZE value, where $N$ is the number of message bits and ZE can be calculated by Eq. 14:

$$ZE = \frac{ZCC_{voiced}}{STE_{voiced}} \qquad (14)$$

4- Save indices of the selected frames as *key*.
5- Embedding every single bit of the message into one selected frame, *f*, as follows:
   a. **DWT:** applying a 2-level discrete wavelet transform on *f*, and selecting the approximation parts in each level for embedding. This step helps the frame being more correlated. So, in the GBT step, it can have a better de-correlation.
   b. **GBT:** applying Graph-based Transform on the approximation coefficients after the 2-level DWT.
   c. **SVD:** applying Singular Value Decomposition on the determined GBT coefficients.

d. **Embed:** embedding one bit of the message to the biggest singular value ($S_{max}$) according to $S$ matrix after the SVD. The embedding can be done by Eq. 15:

$$S'_{max} = \begin{cases} S + \alpha, & \text{if message bit} = 1 \\ S - \alpha, & \text{if message bit} = 0 \end{cases}, \quad \alpha > 0 \quad (15)$$

e. **Save** $S_{max}$ for the extraction.
6- Inverse of SVD by S'$_{max}$.
7- Inverse of GBT.
8- Inverse of 2-level DWT.
9- Attaching all frames again and creating the Stego audio signal.

The extraction can be done as follows:
1- Dividing the stego signal into frames similar to the embedding process.
2- Selecting stego frames based on the *key*.
3- Applying DWT, GBT, and SVD similar to the embedding process for each frame to determine S'$_{max}$.
4- Comparing the determined S'$_{max}$ with the saved S$_{max}$ during the embedding process for each frame.
5- Extract the message bits in each frame using S'$_{max}$ and S$_{max}$ according to Eq. 16.

$$\text{message bit} = \begin{cases} 1, & \text{if } S'_{max} > S_{max} \\ 0, & \text{otherwise.} \end{cases} \quad (16)$$

To use the GBT efficiently, we should define a proper structure for the graph. A good graph should show the correlations between the audio samples. A proper graph structure has been introduced in [1, 2]. In this structure, each audio sample has a link with two nearest neighbors which means each sample is highly correlated with its neighbors. Since we use DWT and also voiced frames, the GBT input is a low-frequency signal. So, it guarantees the correlation and similarity between near sample neighbors. Therefore, the proposed graph structure by Farzaneh, et al [1, 2] which has been shown in Fig.2, is a good structure for our proposed method, too.

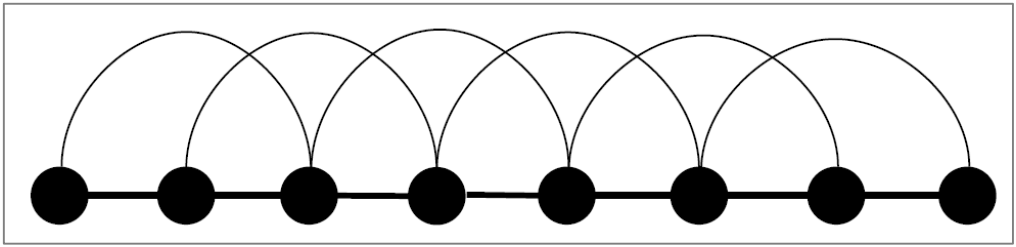

*Figure 2. The proposed graph structure for GBT*

## 5. Experimental Results

We have implemented the proposed method in MATLAB 2018b and Windows 10 OS. The proposed method has tested for 30 speech signals in the NOIZEUS speech database [11]. There are 15 men speech and 15 women speech in this database and each signal has a duration of two seconds. Every signal in NOIZEUS is an English sentence and the sampling rate for all signals is equal to 8 kHz.

To evaluate the imperceptibility we have used Peak Signal to Noise Ratio (PSNR), Short-Time Objective Intelligibility (STOI) [12], and Perceptual Evaluation of Speech Quality (PESQ) [13] measure values. STOI is a real value between zero and one which should be maximized to get good intelligibility of the speech. PESQ is another measurement to evaluate speech quality which is a score between -0.5 to 4.5 and should be maximized to get a high-quality speech signal. To calculate PSNR, PESQ, and STOI, we have compared the original (cover) signals and stego signals (cover signal after embedding the message).

For all experiments, we have used a 50-bit secret message and α value in Eq. 15 has set to 0.05. The frame length is equal to 10 ms (80 samples per frame). So, after a 2-level DWT, we have a low-frequency input signal for GBT with a length of 20 coefficients. After the GBT we use the 16 first coefficients to create a 4×4 matrix and then apply SVD.

The adjacency matrix for graph structure in GBT is a 20×20 matrix which has a value of 1 for first neighbors, and 0.3 for second neighbors.

Table 1 shows the imperceptibility evaluation results based on average values of PSNR, STOI, and PESQ for NOIZEUS signals. Also, Fig.3 shows a sample signal before and after the steganography. As can be seen in Fig.3 there is no visually perceptible change in the stego signal. Also, numerical results are high enough. So, we can say that the proposed method is imperceptible.

*Table 1. Average imperceptibility evaluation results between the cover and the stego speech signals*

| Method | PSNR | PESQ | STOI |
|---|---|---|---|
| DWT-GBT-SVD (proposed) | 52.29 | 4.063 | 0.9749 |
| GBT-SVD [2] | 43.26 | 3.460 | 0.9300 |

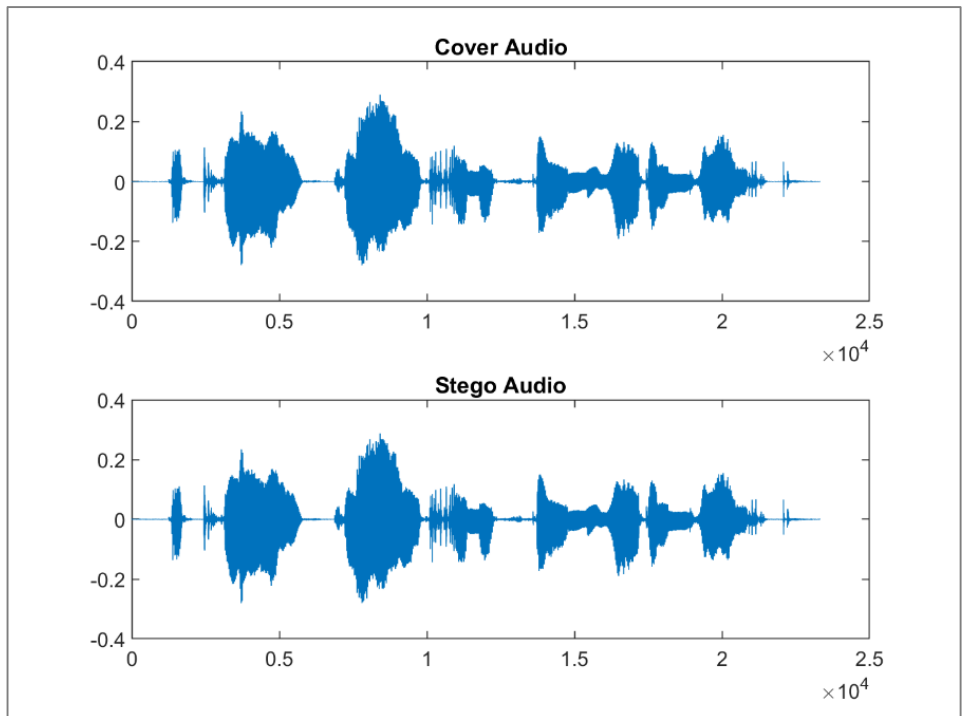

*Figure 3. An example for comparing cover and stego signals*

To evaluate the robustness of the proposed steganography method we have applied different attacks on stego signals before the extraction. The extracted binary message has been compared with the original message and the average Bit Error Ratio (BER) is reported for the NOIZEUS database in Table 2. Also, BER results have been compared with other related studies.

*Table 2. Comparison average BER with other related methods for NOIZEUS database*

| Attack | DWT-GBT-SVD (proposed) | GBT-SVD [2] | DCT-SVD [14] | DWT-SVD [15] | DWT-LU [16] | DWT-DCT [17] |
|---|---|---|---|---|---|---|
| Additive Gaussian Noise-20dB | 0 | 0 | 0 | 0 | - | 0.030 |
| MP3 | 0.210 | 0.232 | 0.001 | 0.290 | 0.020 | 0.010 |
| Resampling | 0 | 0 | 0.030 | 0 | 0 | 0.090 |
| Low Pass Filter (4 kHz) | 0 | 0 | 0 | 0.189 | 0.030 | 0.040 |
| High Pass Filter (50 Hz) | 0 | 0.060 | 0.406 | 0.358 | - | - |
| Amplitude Scaling (0.7) | 0.518 | 0.548 | 0.31 | - | - | 0 |
| Re-quantization | 0 | 0 | 0 | 0 | 0 | 0 |

As we can see in Table 2, the proposed method is robust against lots of attacks, and in comparison to GBT-SVD [2] method, it is completely more robust and more imperceptible. But the proposed method has a big BER for MP3 and Amplitude Scaling attacks. This is because of the α value being small. By increasing the value of α, the robustness will be improved, but the imperceptibility will be decreased. So, we need to find the optimal value for α to make a good trade-off between robustness and imperceptibility.

Fig. 4 shows the STOI and BER for different values of α when the stego signals are under a scaling attack with 0.7 amplitude. As we can see, by increasing the α, robustness has increased (BER has decreased), but imperceptibility has decreased (STOI has decreased). So, we can make the method robust by choosing big values for α, but the method would be perceptible, which is not good for a steganography application. But it could be a good choice for watermarking applications. To get the most imperceptible steganography, we should set α=0.01, and to get the most robust steganography we should set α=0.35. But in this case, the optimal value for α is 0.2. Because the BER is about 0.1 and STOI is big enough.

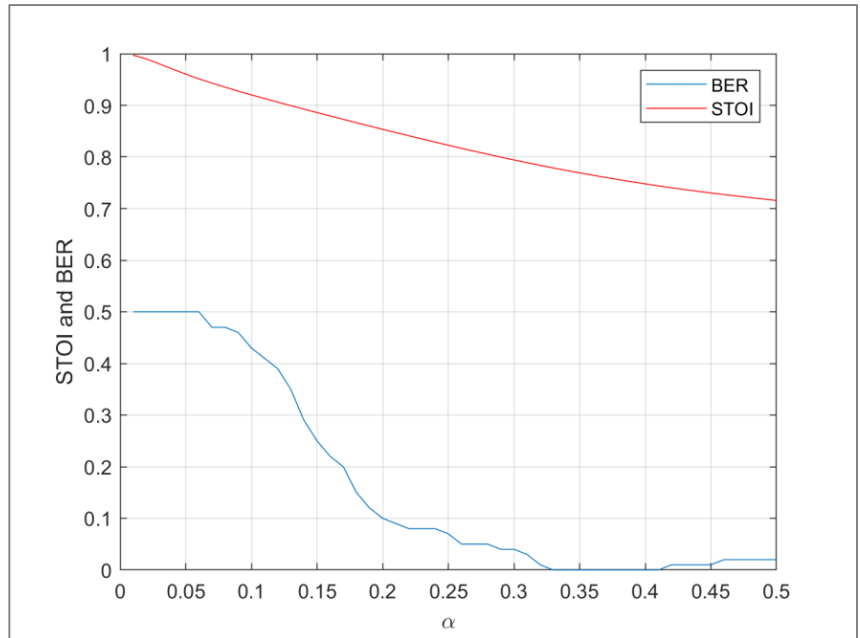
*Figure 4. α value effect on imperceptibility and robustness*

## 6. Conclusion

In this paper, we proposed both a robust and imperceptible speech steganography method that is based on the discrete wavelet transform (DWT), graph-based transform (GBT), and singular value decomposition (SVD). In this method we just used the voiced frames for embedding and results show that the proposed method is completely robust against Gaussian noise, re-sampling, re-quantization, high pass, and low pass filters. Also, it is imperceptible simultaneously. By losing a few imperceptibility we can make the proposed method more robust against MP3 compression and amplitude scaling, too. This shows the importance of finding the watermark strength value or α.

In future researches, we will work on different graph structures for GBT and also different embedding strategies. Also, there are several parameters that can have an effect on imperceptibility and robustness, such as frame length, the number of levels in DWT, etc. In future works, we will find the optimal settings to achieve the best possible steganography.